\documentclass[conference,letterpaper]{IEEEtran}
\usepackage{epsfig,epsf,rotating,setspace,latexsym,amsmath,amssymb,amsfonts,bm,theorem,subfigure,epstopdf,cite,authblk,bbm}
\usepackage{algorithm}
\usepackage[noend]{algpseudocode}
\usepackage{color}
\usepackage{mathtools}
\usepackage{multirow}
\usepackage{soul}

\algrenewcommand\algorithmicforall{\textbf{foreach}}
\algrenewcommand\algorithmicindent{.8em}

\newtheorem{remark}{Remark}

\IEEEoverridecommandlockouts
\allowdisplaybreaks

\begin{document}
\title{Private Counterfactual Retrieval With \\ Immutable Features}

\author{Shreya Meel \quad Pasan Dissanayake \quad Mohamed Nomeir \quad Sanghamitra Dutta \quad Sennur Ulukus \\
\normalsize Department of Electrical and Computer Engineering \\
\normalsize University of Maryland, College Park, MD 20742 \\
\normalsize {\it smeel@umd.edu} \quad {\it pasand@umd.edu} \quad{\it mnomeir@umd.edu} \quad {\it sanghamd@umd.edu} \quad {\it ulukus@umd.edu}}

\maketitle

\begin{abstract}
In a classification task, counterfactual explanations provide the minimum change needed for an input to be classified into a favorable class. We consider the problem of privately retrieving the exact closest counterfactual from a database of accepted samples while enforcing that certain features of the input sample cannot be changed, i.e., they are \emph{immutable}. An applicant (user) whose feature vector is rejected by a machine learning model wants to retrieve the sample closest to them in the database without altering a private subset of their features, which constitutes the immutable set. While doing this, the user should keep their feature vector, immutable set and the resulting counterfactual index information-theoretically private from the institution. We refer to this as immutable private counterfactual retrieval (I-PCR) problem which generalizes PCR to a more practical setting. In this paper, we propose two I-PCR schemes by leveraging techniques from private information retrieval (PIR) and characterize their communication costs.  Further, we quantify the information that the user learns about the database and compare it for the proposed schemes.
\end{abstract}

\section{Introduction}
%counterfactuals important for explainability
The right to explanations mandates that any black box machine learning model making crucial decisions in high-stakes applications should provide the user, i.e., the applicant, with a suitable explanation for its decision~\cite{voigt2017eu}. In this regard, counterfactual explanations have grown as an effective means to deliver the minimum perturbation required at the user's end to alter the model's decision to a favorable outcome \cite{Harvard_discussion_first_counterfactual}. For instance, in the case of a bank loan rejection, a user might receive a counterfactual recommendation to increase their income by $10$K to get accepted. Numerous works have focused on generating counterfactuals with different properties, namely, proximity to the user's input \cite{nice}, robustness to model changes \cite{upadhyayROAR, hammanRobustCF, dutta2022robust, hammanRobustCFJournal}, feasibility under user's constraints \cite{face, dice}, sparsity in the attributes to be changed \cite{nice,dice}, and diversity of the counterfactuals \cite{dice}. We refer the reader to \cite{verma2020counterfactual, guidottiCounterfactualSurvey, mishra2021survey} for a comprehensive survey on different methods.

%privacy problem and the research gap of user's privacy concern
Providing an appropriate counterfactual poses serious privacy concerns, both for the user asking for a counterfactual and for the institution delivering it. Existing works such as  \cite{pawelczykMembershipInference, yang2022differentially, privacy_issue_in_cf, pentyala2023privacy} focus on preserving data privacy from the institution's side, while \cite{aivodjiModelExtraction, wangDualCFModelExtraction, dissanayakeModelExtraction} focus on the extraction of the model by querying for multiple counterfactual explanations. However, preserving privacy from the user's side has rarely received attention. We are particularly interested in the scenario where a user would like to obtain a counterfactual explanation without revealing their input feature vector to the institution. The user may be reluctant to share their feature vector with the institution for several reasons, e.g., if they have a limited number of attempts to apply, or if they wish to preserve their data privacy until they improve their chances of acceptance. 

\begin{figure}[t]
    \centering
    \includegraphics[width=0.78\linewidth]{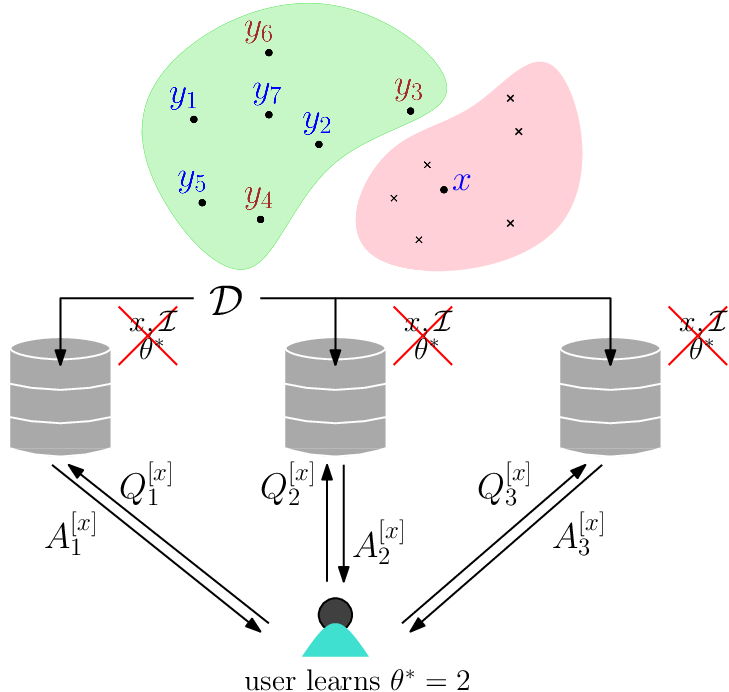}
    \caption{System model where $y_1$, $y_2$, $y_5$, $y_7$ have the same values for the immutable features as $x$, i.e., are viable counterfactuals for $x$.}
    \label{sysmod_immutable}
    \vspace*{-0.3cm}
\end{figure}

%closely related work on ann's, our previous work and where they fall short, survey on pir
The first work that focuses on user's privacy in retrieving counterfactual explanations is \cite{nomeir2024privatecounterfactualretrieval} where the private counterfactual retrieval (PCR) problem is introduced and formulated. This work leverages techniques from private information retrieval (PIR), which is a subject of independent interest, to obtain the index of the nearest counterfactual. In PIR \cite{chor}, a user wishes to retrieve one message out of $K$ replicated messages in $N$ servers without leaking any information about their required message index. The capacity of PIR, i.e., the maximum ratio between the number of the required message symbols and the total downloaded symbols, is found in \cite{c_pir}. In symmetric PIR (SPIR), an extra requirement is considered where the user cannot get any information about message symbols aside from their required message. The capacity for such a model is found in \cite{c_spir}. Other important variants can be found in \cite{arbitrarycollusion, banawan_eaves, banawan_multimessage_pir, banawan_pir_mdscoded, batuhan_hetero, byzantine_tpir, C_SETPIR, ChaoTian, codedstorage_adversary_tpir, colluding, csa, first_xsecure, wei_banawan_cache_pir, wang_spir, ulukusPIRLC, uncoded_constrainedstorage_pir, tspir_mdscoded, utah_hetero, tpir_sideinfo, sun_eaves, semantic_pir, Salim_CodedPIR, nan_eaves}. Notably, the PCR strategy proposed in \cite{nomeir2024privatecounterfactualretrieval} allows all features of the user's input to be altered while obtaining a counterfactual on the accepted side. However, this assumption is unsuitable for practical settings where certain attributes (e.g., nationality, place of birth, gender) are immutable for a user, restricting the set of attainable counterfactuals.

%our contributions in this paper 
In this paper, we allow the user to fix a subset of their features as \emph{immutable}, i.e., their values should not be altered in their corresponding counterfactual. The immutable set is user-specific and should be kept private from the institution. With this additional constraint, we propose immutable PCR (I-PCR) wherein the user retrieves their counterfactual index from a database of accepted feature vectors. To obtain the counterfactual vector, the user simply runs an SPIR protocol with this index after I-PCR. We present two achievable schemes, namely, a two-phase and a single-phase I-PCR, and compare their communication and computation costs through runtime analyses on real datasets. Although these schemes guarantee information theoretic privacy for the user, the leakage from the database is non-zero. We find that the single-phase scheme incurs lower communication cost than the two-phase one, while causing more database leakage. We evaluate this leakage numerically for both schemes.  Further, we account for \emph{user actionability} which guarantees preferential change of certain attributes compared to the rest. We demonstrate that our schemes can be modified to include private user actionability on the mutable attributes without compromising the privacy of the immutable attributes. 

\section{System Model}
We assume a pre-trained two-class classification model that takes as input a $d$-dimensional feature vector and classifies it into its target class, e.g., accepted or rejected. A user who is rejected by this model wishes to privately retrieve a valid counterfactual sample corresponding to their data sample while keeping some of their features fixed. The user does not have access to the model and relies on a database $\mathcal{D}$ that contains the feature vectors of a set of samples accepted by the same model. We assume that each attribute of the samples is an integer in $[0:R]=\{0, 1, \ldots, R\}$. The samples in $\mathcal{D}$ are indexed as $y_1, y_2, \ldots, y_M$ where $M=|\mathcal{D}|$ and are stored in $N$ non-colluding and non-communicating servers in a replicated manner. The system model is illustrated in Fig.~\ref{sysmod_immutable}.

Let $x\in [0:R]^d$ denote the user's sample. The goal of the user is to retrieve the index of their counterfactual, i.e., the sample in $\mathcal{D}$ which is closest in terms of a distance metric $d(\cdot,\cdot)$ to $x$. Moreover, this nearest sample should differ from $x$ in at most a subset $\mathcal{M}\in [d]$ of features. Alternatively, the user fixes their counterfactual's feature values in $\mathcal{I} = [d]\setminus \mathcal{M}$, which we call the \emph{immutable set}. In particular, let $\mathcal{I}=\{j_1, j_2, \ldots, j_{|\mathcal{I}|}\}$ with $j_1<j_2<\ldots<j_{|\mathcal{I}|}$, then 
\begin{align}
    x_{\mathcal{I}} = 
    \begin{bmatrix}
        x_{j_1} & x_{j_2} & \ldots & x_{j_{|\mathcal{I}|}}
    \end{bmatrix}^T,
\end{align} 
and $x_{\mathcal{M}}$ is defined in a similar way. Therefore, the goal of the user is to find the counterfactual index $\theta^*$ given by
\begin{align}
    \theta^*=\arg \min_{i\in \Theta} d(x,y_i)
    \text{ with }\Theta = \{i\in [M]: y_{i,\mathcal{I}}=x_{\mathcal{I}}\},
\end{align}
where the knowledge of $\mathcal{I}$, hence $\Theta$, is restricted to the user and no information on $x$ is revealed to the server. We use the squared $\ell_2$ norm as the distance metric, $d(x,y_i) = ||y_i-x||^2$. 

To this end, the user sends to each server $n\in [N]$ the query $Q_n^{[x]}$. Upon receiving the query, each server computes $A_n^{[x]}$, i.e., their answers using their storage, the query and the shared common randomness $Z'$, i.e.,
\begin{align}
    H(A_n^{[x]}|\mathcal{D},Q_n^{[x]},Z') = 0.
\end{align}
The common randomness $Z'$ is shared only among the servers before initiating the I-PCR scheme and is independent of $\mathcal{D}$. Using the responses from all of the servers, the user determines the index $\theta^*$ of their corresponding counterfactual, i.e., 
\begin{align}\label{eq_decodability}
    [\text{Decodability}] \quad H(\theta^*| Q_{[N]}^{[x]}, A_{[N]}^{[x]}, x,\mathcal{I}) = 0.
\end{align}
To achieve user privacy, each server should not be able to acquire any information on the user's sample, their immutable set or the index of their counterfactual, i.e., $\forall n \in [N]$,
\begin{align}\label{eq_userprivacy}
   [\text{User Privacy}] \quad I(x, \theta^*, \mathcal{I}; Q_n^{[x]}, A_n^{[x]}| \mathcal{D})=0.
\end{align}
On the other hand, the answers from the servers reveal some information on $\mathcal{D}$ to the user. This leakage is quantified as,
\begin{align}\label{eq_leakage_main}
[\text{Database Leakage}] \quad I(y_{[M]}; Q_{[N]}^{[x]}|x, \mathcal{I}).
 \end{align}

An I-PCR scheme is said to be \emph{achievable} if it simultaneously satisfies \eqref{eq_decodability} and \eqref{eq_userprivacy}. Its communication cost is given by the total number of symbols transmitted between the user and the servers while executing the scheme.

\section{Achievable Schemes}
In this section, we describe two achievable I-PCR schemes each requiring $N=3$ replicated servers storing $\mathcal{D}$.

\subsection{Two-Phase I-PCR}
The scheme proceeds in two phases. In the first phase, the user retrieves $\Theta$, i.e., the indices $i$ of $y_i$ for which ${x}_{\mathcal{I}} = y_{i,\mathcal{I}}$. In the second phase, the user retrieves the exact squared distances of the samples indexed by $\Theta$ and compares them to determine the counterfactual index $\theta^*\in \Theta$. The scheme operates in a field $\mathbb{F}_q$ where $q>R^2d$ is a prime. Further, the set of queries and answers for the $n$th server over two phases  are $Q_n^{[x]}=[Q_n^{[x_{\mathcal{I}}]}, Q_n^{[x_{\mathcal{M}}]}]$ and $A_n^{[x]}=[A_n^{[x_{\mathcal{I}}]}, A_n^{[x_{\mathcal{M}}]}]$, respectively.

\paragraph{Phase 1} Based on the user's choice of $\mathcal{I}$, the user constructs the binary vector $h_1\in \{0,1\}^{d}$ such that,
\begin{align}
    h_1(k) = \mathbbm{1}\left[k \in \mathcal{I}\right], \quad k\in [d].
\end{align}
Then, the user sends the following query tuple to server $n$,
\begin{align}
    Q_n^{[x_{\mathcal{I}}]} = &\big[ Q_n^{[x_{\mathcal{I}}]}(1), Q_n^{[x_{\mathcal{I}}]}(2)\big] \\ =&\big[ h_1+\alpha_n Z_1, x\circ h_1 + \alpha_n Z_2 \big],
\end{align}
where $\circ$ denotes element-wise product, and $Z_1, Z_2$ are chosen uniformly and independently from $\mathbb{F}_q^{d}$ by the user. Upon receiving the query, server $n$ responds with the following answer corresponding to each $i\in [M]$,
\begin{align}
    A_n^{[x_{\mathcal{I}}]}(i) =& \rho_i|| Q_n^{[x_{\mathcal{I}}]}(1)\circ y_i - Q_n^{[x_{\mathcal{I}}]}(2)||^2 + \alpha_n Z_1'(i)\notag \\
    &+ \alpha_n^2 Z_2'(i),
\end{align}
where $\rho_i$ is chosen uniformly at random from $[1:q-1]$ by the servers, $Z_1'(i)$ and $Z_2'(i)$ are uniform random variables from $\mathbb{F}_q$. Thus,
\begin{align}
     A_n^{[x_{\mathcal{I}}]}(i) =& \rho_i || (h_1+\alpha_n Z_1)\circ y_i - (x\circ h_1 + \alpha_n Z_2)||^2\notag \label{answer_twophase}\\
     &+ \alpha_n Z_1'(i)+ \alpha_n^2 Z_2'(i)\\
     =& \rho_i ||h_1\circ (y_i - x)||^2 \nonumber \\ 
     &+ \alpha_n \big(Z'_1(i)+2 \rho_i (h_1\circ (y_i - x))^T(Z_1\circ y_i-Z_2)\big)\notag \\
     &+ \alpha_n ^2 \big( \rho_i||Z_1\circ y_i - Z_2||^2+ Z'_2(i)\big).
\end{align}
From the answers of the three servers, the user obtains the answer vector $A^{[x_{\mathcal{I}}]}(i)$ as
\begin{align}\label{answer_phase1}
    A^{[x_{\mathcal{I}}]}(i) = 
    \begin{bmatrix}
        A_1^{[x_{\mathcal{I}}]}(i)\\
        A_2^{[x_{\mathcal{I}}]}(i)\\
        A_3^{[x_{\mathcal{I}}]}(i)
    \end{bmatrix}
    =
    \bm{M}_3 
    \begin{bmatrix}
        \rho_i||h_1\circ (y_i - x)||^2 \\
        I_{11}(i)\\
        I_{12}(i)
    \end{bmatrix},
\end{align}
where
\begin{align}
    \bm{M}_3 = \begin{bmatrix}
        1 & \alpha_1 & \alpha_1^2 \\
        1 & \alpha_2 & \alpha_2^2\\
        1 & \alpha_3 & \alpha_3^2
    \end{bmatrix},
\end{align}
is a Vandermonde matrix of dimension $3$, hence is invertible when $\alpha_1, \alpha_2$ and $\alpha_3$ are distinct elements from $\mathbb{F}_q$, and $I_{11}(i)$, $I_{12}(i)$ are interference terms that reveal no information on $\mathcal{D}$ to the user. For each $i$, the user concludes that, $i \in \Theta$ iff
\begin{align}
 \rho_i||h_1\circ (y_i - x)||^2 = 0.
\end{align}

A non-zero value of  $\rho_i||h_1\circ (y_i - x)||^2$ implies that $y_{i,\mathcal{I}}\neq x_{\mathcal{I}}$ since $\rho_i$ is a non-zero element of $\mathbb{F}_q$. Further, since $\rho_i$s are independent random variables, the user does not learn which samples in $\mathcal{D}$ share some immutable attributes, unless they are identical to $x_{\mathcal{I}}$. If $\Theta=\emptyset$, then the user learns that there are no valid counterfactuals in $\mathcal{D}$ satisfying their current requirement, and the scheme ends here. Further, if $|\Theta|=1$, the user does not need the second phase.

\paragraph{Phase 2} The user now compares the distances in a subset of $\mathcal{D}$, i.e., $\{y_i: i\in \Theta\}$ to find their counterfactual index $\theta^*$ without revealing $\Theta$ to the servers. Towards this, the user prepares the following binary vector $h_2 \in \{0,1\}^{M}$,
\begin{align}
    h_2 (i) = 
    \begin{cases}
        1, & \text{ if $i \in \Theta$},\\
        0, & \text { otherwise.}
    \end{cases}
\end{align}
Then, the user sends the following query tuple to server $n$,
\begin{align}
    Q_n^{[x_{\mathcal{M}}]} = & [Q_n^{[x_{\mathcal{M}}]}(1),  Q_n^{[x_{\mathcal{M}}]}(2)] \\ = & [h_2 + \alpha_n Z_3, x + \alpha_n Z_4],
\end{align}
where $Z_3$ and $Z_4$ are uniform vectors from $\mathbb{F}_{q}^{M}$ and $\mathbb{F}_{q}^{d}$, respectively. Each server constructs their masked database $\tilde{S}_n$ as follows,
\begin{align}
     \tilde{S}_n = \text{diag}(Q_n^{[x_{\mathcal{M}}]}(1)) \times \begin{bmatrix}         y_1 & y_2 & \ldots & y_M
     \end{bmatrix}^T.
 \end{align}

Let $\tilde{S}_n(i)$ denote the $i$th row of the masked database at server $n$. Then, their corresponding answer is given by,
\begin{align}
    A_n^{[x_{\mathcal{M}}]} (i) = &||\tilde{S}_n (i)^T - Q_n^{[x_{\mathcal{M}}]}(2)||^2 + \alpha_n Z'_3(i) + \alpha_n^2 Z'_4(i)\\
    =& ||(h_2(i) + \alpha_n Z_3(i)) y_{i} - (x+ \alpha_n Z_4)||^2\notag \\
    &+ \alpha_n Z'_3(i) + \alpha_n^2 Z'_4(i)\\
    =& ||h_2(i) y_{i} - x||^2 \nonumber\\
    &+ \alpha_n \big(2(h_2(i)y_{i} -x)^T(Z_3(i) y_{i} - Z_4) + Z'_3(i)\big) \notag \\
    &+ \alpha_n^2\big(||Z_3(i)y_{i} - Z_4||^2 + Z'_4(i)\big),
\end{align}
where $Z_3'(i)$ and $Z_4'(i)$ are uniform random variables from $\mathbb{F}_q$. From the answers of all three servers, the user computes the answer vector,
\begin{align}\label{answer_phase2}
    A^{[x_{\mathcal{M}}]} (i) = \bm{M}_3
    \begin{bmatrix}
        ||h_2(i) y_{i} - x||^2 \\
        I_{21}(i)\\
        I_{22}(i)
    \end{bmatrix},
\end{align}
where the first term reduces to
\begin{align}
 ||h_2(i) y_{i} - x||^2=
 \begin{cases}
     ||y_{i} - x||^2=||y_{\mathcal{M},i} - x_{\mathcal{M}}||^2, & i\in \Theta,\\
     ||x||^2, & \text{o.w.}
 \end{cases}
\end{align} 
In other words, if the values of the immutable features do not match, the user does not learn any information (not even their distances) on the samples $y_i$, $i\notin \Theta$. 

\begin{remark}
    The above scheme requires $N=3$ servers, and incurs an upload cost of $9d+3M$ and a download cost of $6M$. Instead of sending $Q_n^{[x_{\mathcal{I}}]}$ and $Q_n^{[x_{\mathcal{M}}]}(2)$, the user can ask the servers to compute $x\circ h_1$ by securely sending $x$ and $h_1$ in the query tuple, $Q_n^{[x_{\mathcal{I}}]}=[h_1+\alpha_n Z_1, x+\alpha_n Z_2]$. Server $n$ computes the product $(h_1+\alpha_n Z_1)\circ (x+\alpha_n Z_2) = x\circ h_1 + \alpha_n \hat{Z}_1 + \alpha_n^2 \hat{Z}_2$ where $\hat{Z}_1$ and $\hat{Z}_2$ are the resulting interference terms. Moreover, the first term in the answer \eqref{answer_twophase} is $\rho_i || (h_1+\alpha_n Z_1)\circ y_1 - (x\circ h_1 + \alpha_n \hat{Z}_1 + \alpha_n^2 \hat{Z}_2)||^2$ (added to $\alpha_n Z'_1(i)+ \alpha_n^2 Z'_2(i) + \alpha_n^3 Z'_3(i)$) leading to a polynomial of degree $4$ in $\alpha_n$. However, since $\alpha_n^4||\hat{Z}_2||^2, n\in[4]$ is known to the user, it can be canceled from the answers resulting in a reduced degree of $3$. This requires $N=4$ (instead of $3$) replicated servers to retrieve $\Theta$ and the communication cost in the first phase is $8d+4M$. In the second phase, the user sends a single query $Q_n^{[x_{\mathcal{M}}]}=Q_n^{[x_{\mathcal{M}}]}(1)=h_2+\alpha_n Z_3$ to any $3$ out of the $4$ servers who then compute their answers as described in the aforementioned scheme. This results in an upload cost of $3M$ and download cost of $3M$. The total communication cost is, therefore, $8d+10M$. Although the scheme requires an additional server, its  communication cost is lower than the previous scheme,  if $d>M$ .
\end{remark}

\subsection{Single-Phase I-PCR}
In this section, we assume that the user assigns at most $F$ features as immutable, where $F$ is globally known. Thus, $\mathcal{I}\subset [d]$ where $|\mathcal{I}|\leq F$. We show that the baseline PCR+ scheme of \cite{nomeir2024privatecounterfactualretrieval} can be modified to develop an achievable I-PCR scheme with a single round of queries and answers. To this end, the servers and the user agree on $\mathbb{F}_q$ with $q>F(L-1)R^2+R^2d$ and prime. The user chooses an integer $L>R^2 d$ as a scaling factor for the immutable attributes. In particular, the user designs a weighing vector $h\in \{1,L\}^d$,
\begin{align}\label{weight_assignment}
    h(k) = 
    \begin{cases}
        L, & \text{ if $k\in \mathcal{I}$},\\
        1, & \text{ otherwise.}
    \end{cases}
\end{align}
To server $n$, the user sends the query tuple,
\begin{align}
    Q_n^{[x]}=&[ Q_n^{[x]}(1),  Q_n^{[x]}(2)] \\ =&[x+\alpha_n Z_1, h+\alpha_n Z_2],
\end{align}
where $Z_1$ and $Z_2$ are independent uniform random vectors in $\mathbb{F}_q^d$ that are private to the user. In response, server $n$ computes the following answer corresponding to $y_i$,
\begin{align}
    A_n^{[x]}(i)=& (y_i - Q_n^{[x]}(1))^T\big((y_i - Q_n^{[x]}(1))\circ Q_n^{[x]}(2) \big)\notag \\
    &+\alpha_n Z'_1(i) + \alpha_n^2 Z'_2(i)\\
    =& (y_i-x)^T \big((y_i-x)\circ h\big) + \alpha_n I_1(i) +\alpha_n ^2 I_2(i)\nonumber\\
    &+ \alpha_n^3 Z_1^T(Z_1\circ Z_2),
    % &+ \alpha_n \bigg( (y_i-x)^T\big((y_i-x)\circ Z_2\big)- 2(y_i-x)^T(Z_1 \circ h) +Z'_1(i)\bigg)\nonumber\\
    % &+ \alpha_n^2 \big( Z_1^T \bm{H}Z_1 - 2Z_1^T \bm{Z}(y_i-x)  +Z'_2(i) \big) + \alpha_n^3 Z_1^T\bm{Z}Z_1,
\end{align}
where
\begin{align}
    I_1(i) =&(y_i\!-\!x)^T\big((y_i\!-\!x)\circ Z_2\big)- 2(y_i\!-\!x)^T(Z_1 \circ h\big) +Z'_1(i)\\
    I_2(i) =& Z_1^T\big(Z_1\circ h\big) - 2 Z_1^T\big((y_i\!-\!x)\circ Z_2\big)+Z_2'(i).
\end{align} Upon receiving the answers from 3 servers, the user cancels the term of third degree, since it is known to them, i.e.,
\begin{align}
    \hat{A}_n(i) = A_n^{[x]}(i) - \alpha_n^3  Z_1^T(Z_1\circ Z_2),
\end{align}
which gives
\begin{align}\label{answer_single_phase}
     \hat{A}(i) = \begin{bmatrix}
        \hat{A}_1(i)\\
        \hat{A}_2(i)\\
        \hat{A}_3(i)
    \end{bmatrix} = \bm{M}_3\begin{bmatrix}
        (y_i-x)^T \big((y_i-x)\circ h\big)\\
        I_1(i)\\
        I_2(i)
    \end{bmatrix}.
\end{align}
Now, to determine $\Theta$, the user checks the range of $(y_i-x)^T \big((y_i-x)\circ h\big)$. If it lies in $[0:(d-|\mathcal{I}|)R^2]$, all immutable features match in value. Among these, the user selects the index with the minimum value of $(y_i-x)^T \big((y_i-x)\circ h\big)$ as $\theta^*$. On the other hand, if at least one immutable feature does not match, the resulting range of $(y_i-x)^T \big((y_i-x)\circ h\big)$ is $[L: L|\mathcal{I}|R^2 + (d-|\mathcal{I}|)R^2]$. Since $L$ is chosen such that $L>R^2d$, these two sets are disjoint and the user is able to distinguish between these samples.

\begin{remark}
    Note that the distance metric in the system model and proposed schemes uses the $\ell_2$ norm. The system model and the schemes can be extended to any $\ell_k$ norm, where $k$ is even by modifying the answer generation accordingly. 
\end{remark}

\section{Comparison of the Above Schemes} 
Given the aforementioned I-PCR achievable schemes, we compare their performance based on the following metrics.

\subsection{Communication Cost} 
For both of the schemes, $N=3$ non-colluding servers with replicated $\mathcal{D}$ are required. For the two-phase scheme, the communication cost is $9(d+M)$, whereas, for the single-phase scheme, the upload cost is $2d$ symbols to each server, while the download cost is $M$ symbols from each server. This results in a communication cost of $6d+3M$. The single-phase I-PCR scheme saves on the communication cost by $3d+6M$.

\subsection{Field Size Requirement} 
In order to ensure decodability, the choice of the field size is important. For the two-phase scheme, the prime $q$ should be greater than $R^2 d$. For the single-phase scheme, the field size should satisfy $q>L|\mathcal{I}|R^2+R^2(d-|\mathcal{I}|)=|\mathcal{I}|(L-1)R^2+R^2d$ for a given $|\mathcal{I}|$. This requires $q>F(L-1) R^2+R^2d$, where $F$ is the upper bound on $|\mathcal{I}|$. Note that no such upper bound is required for the two-phase scheme. 

\subsection{Incorporating User Actionability} Now, assume that, among the features in $\mathcal{M}$, the user has a higher preference/willingness to change certain attributes over others. To realize this, the user assigns an integer $w'(k)\in [L_1]$ for each $k\in \mathcal{M}$ where a higher value indicates more reluctance to change that attribute and vice versa. Then, the user's goal is to minimize:
\begin{align}
    (y_{i,\mathcal{M}}-x_{\mathcal{M}})^T \big((y_{i,\mathcal{M}}-x_{\mathcal{M}})\circ [w'(k)]_{k\in \mathcal{M}}\big)
\end{align}
over all $y_i\in \mathcal{D}$ such that $y_{i,\mathcal{I}}=x_{\mathcal{I}}$. The aforementioned schemes can be modified to achieve this additional requirement as described next.

For the two-phase I-PCR scheme, the field size requirement is now increased to $q>L_1R^2d$. The first phase remains the same requiring $N=3$ replicated servers while the second phase requires an additional server, i.e., $N=4$. In the second phase, the user sends $w+\alpha_n Z_5\in \mathbb{F}_q^d$ as their third query $Q_n^{[x_{\mathcal{M}}]}(3)$ to server $n\in [4]$, where
\begin{align}
    w(k)=\begin{cases}
        1, & k\in \mathcal{I},\\
        w'(k), & k\in \mathcal{M}.
    \end{cases}
\end{align}
The first term in the answer changes to $\big(\tilde{S}_n (i)^T - Q_n^{[x_{\mathcal{M}}]}(2)\big)^T\big(\tilde{S}_n (i)^T - Q_n^{[x_{\mathcal{M}}]}(2) \circ (w+\alpha_n Z_5)\big)$ and another randomness term $\alpha_n^3 Z'_5(i)$ is added. The second phase requires $4$ servers for decodability and the total communication cost incurred is $14d+11M$ symbols.

With the single-phase scheme, actionability can be directly implemented by modifying \eqref{weight_assignment} as follows,
\begin{align}
    h(k) = \begin{cases}
        L, & k\in \mathcal{I},\\
        w'(k), & k\in \mathcal{M},
    \end{cases}
\end{align}
where $L>L_1R^2d$. The rest of the scheme follows with $3$ servers leading to the same communication cost $6d+3M$, however, in a larger field $\mathbb{F}_q$, with $q>F(L-1)R^2 + L_1R^2d$.

\subsection{Database Leakage} 
Note that the database leakage, defined in \eqref{eq_leakage_main}, can be rewritten as follows,
\begin{align}
 &I(y_{[M]}; Q_{[N]}^{[x]}, A_{[N]}^{[x]}|x, \mathcal{I}) \nonumber\\
 &\quad=I(y_{[M]}; A_{[N]}^{[x]}|x, \mathcal{I}, Q_{[N]}^{[x]})+ I(y_{[M]}; Q_{[N]}^{[x]}|x, \mathcal{I}) \\
 &\quad=I(y_{[M]}; A_{[N]}^{[x]}|x, \mathcal{I}, Q_{[N]}^{[x]})\label{eq_leakage_indep}
\end{align}
where \eqref{eq_leakage_indep} follows since $y_{[M]}=(y_1,\ldots,y_M)$ are independent of the queries $Q_{[N]}^{[x]}$ given $x$ and $\mathcal{I}$. 

In addition, it is clear that for both schemes, we have,
\begin{align}\label{eq_dropping_Q}
I(y_{[M]}; A_{[N]}^{[x]}|x, \mathcal{I}, Q_{[N]}^{[x]}) = I(y_{[M]}; A_{[N]}^{[x]}|x, \mathcal{I}).
\end{align}
This is because, the randomness symbols in $Q_{[N]}^{[x]}$ added for the privacy of $x,\mathcal{I}$, upon answer generation are masked by the randomness shared among the servers and align on a subspace different from the database information subspace. In particular database leakage for the two-phase scheme originates from $\rho_i||h_1\circ(y_i-x)||^2$ in \eqref{answer_phase1} and $||h_2\circ y_i-x||^2$ in \eqref{answer_phase2}. Similarly, $(y_i-x)^T\big((y_i-x)\circ h\big)$ in \eqref{answer_single_phase} is responsible for leakage in the single-phase scheme.

For comparing the database leakage between the two schemes, define for each $i\in [M]$,
\begin{align}
    E_i = \mathbbm{1}\{x_{\mathcal{I}}=y_{i,\mathcal{I}}\}.
\end{align}
Thus, the answers for both schemes satisfy
\begin{align}\label{eq_E_funcA}
     H(E_{[M]}|A_{[N]}^{[x]}, x, \mathcal{I})&=0,
\end{align}
where $E_{[M]}=(E_1,\ldots,E_M)$. The database leakage 
\begin{align}\label{eq_E_y_joint}
    &I(y_{[M]}; A_{[N]}^{[x]}|x,\mathcal{I})=I(E_{[M]},y_{[M]}; A_{[N]}^{[x]}|x, \mathcal{I}) 
\end{align}
since $I(E_{[M]};A_{[N]}^{[x]}|x,\mathcal{I}, y_{[M]})=0$. The right hand side of \eqref{eq_E_y_joint} can be expanded as
\begin{align}
     &I(E_{[M]};A_{[N]}^{[x]}|x,\mathcal{I})+ I(y_{[M]};A_{[N]}^{[x]}|x,\mathcal{I}, E_{[M]})\nonumber\\
     &\quad= H(E_{[M]}|x,\mathcal{I}) + I(y_{[M]};A_{[N]}^{[x]}|x,\mathcal{I},E_{[M]}), \label{eq_leakage_compare}
\end{align}
where \eqref{eq_leakage_compare} follows from \eqref{eq_E_funcA}. The first term in the leakage appears for both schemes since $\Theta$ is inevitably learned by the user while computing $\theta^*$. The second term, however, is different for the two schemes.

For the two-phase scheme, the second term is the information that the user infers from the squared norm values $||y_i-x||^2=||y_{i,\mathcal{M}}-x_{\mathcal{M}}||^2$ only for $E_i=1$, i.e., $i\in \Theta$. On the other hand, for the single-phase scheme, the second term contains additional information about the immutable features of $y_i, i\notin \Theta$. In particular, if one immutable feature does not match, then $(y_i-x)^T\big((y_i-x)\circ h\big)\in \mathcal{J}_1=[L:LR^2+R^2(d-|\mathcal{I}|)]$ whereas if two immutable features do not match, then $(y_i-x)^T\big((y_i-x)\circ h\big)\in \mathcal{J}_2=[2L:2LR^2+R^2(d-|\mathcal{I}|)]$ and so on. In general $\mathcal{J}_k=[kL:kL+R^2(d-|\mathcal{I}|)]$ is the set where $(y_i-x)^T\big((y_i-x)\circ h\big)$ lies if $k$ out of $|\mathcal{I}|$ features do not agree. These sets, though overlapping, leak non-zero amount of information on $y_i, i\notin \Theta$. For instance, if $(y_i-x)^T\big((y_i-x)\circ h\big)\in [L:2L-1]$, the user learns that $x_{\mathcal{I}}=y_{i,\mathcal{I}}$ in all but one feature. 

\section{Numerical Results}
In this section, we numerically analyze the leakage on $\mathcal{D}$ with $M$ fixed. For the two-phase scheme, we simplify the database leakage expression using \eqref{eq_E_y_joint} and \eqref{eq_leakage_compare}, and get,
\begin{align}
    I(y_{[M]}; A_{[N]}^{[x]}|x,\mathcal{I})\!=\!H(E_{[M]}|x,\mathcal{I})\!+\!I(y_{[M]};A_{[N]}^{[x]}|x,\mathcal{I},E_{[M]}).
\end{align}
Expanding the latter term gives,
\begin{align}
    I(&y_{[M]};A_{[N]}^{[x_{\mathcal{I}}]},A_{[N]}^{[x_{\mathcal{M}}]}|x,\mathcal{I},E_{[M]}) \nonumber\\   =&I(y_{[M]};A_{[N]}^{[x_{\mathcal{I}}]}|x,\mathcal{I},E_{[M]})\!+\!I(y_{[M]};A_{[N]}^{[x_{\mathcal{M}}]}|x,\mathcal{I},E_{[M]},A_{[N]}^{[x_{\mathcal{I}}]})
\end{align}
where the first term on the right hand side is $0$ since the answers after the first phase reveal no more information on $y_{[M]}$ beyond $E_{[M]}$. Now, since $H(A_{[N]}^{[x_{\mathcal{M}}]}|x,\mathcal{I},E_{[M]}, A_{[N]}^{[x_{\mathcal{I}}]},y_{[M]})=0$ the second term reduces to
\begin{align}
    H(A_{[N]}^{[x_{\mathcal{M}}]}|x,\mathcal{I}, E_{[M]},A_{[N]}^{[x_{\mathcal{I}}]})=    H(A_{[N]}^{[x_{\mathcal{M}}]}|x,\mathcal{I}, E_{[M]}).
\end{align}
This is because, for each $i\in[M]$, after decoding $A_{[N]}^{[x_{\mathcal{I}}]}(i)$, 
\begin{align}
    \rho_i||y_{i,\mathcal{I}}-x_{\mathcal{I}}||^2=\nu_i E_i,
\end{align}
and $I(A_{[N]}^{[x_{\mathcal{M}}]}; \nu_i E_i, i\in[M] | x, \mathcal{I}, E_{[M]}) = 0$ since $\rho_i$ is chosen at random from $[1:q-1]$ independently and therefore $\nu_i \in [1:q-1]$ is also independent from $y_{[M]}$, $x$, and $Q_n^{[x_{\mathcal{M}}]}$ because the operations are done in a finite field $\mathbb{F}_q$.

%%%%%%% results %%%%
We compute the exact leakage values on a synthetic dataset, setting $R = 3$, $d = 3$, and the database size $M = 3$. The samples are assumed to be uniformly distributed over $[0:R]^d$ space. For each sample $x$, the counterfactuals $y_1, \ldots, y_M$ are assumed to be uniformly distributed over $[0:R]^d \backslash \{x\}$. Moreover, the set of immutable features $\mathcal{I}$ is assumed to be uniformly distributed over $[d]$, while having a fixed cardinality. For the single-phase scheme, $F=d$, $L=R^2d+1=28$ and $q$ is chosen as $757$ which the smallest prime greater than $d(L-1)R^2+R^2d=756$.
Table~\ref{table_leakageValues} presents the computed leakage values under these conditions, with varying $|\mathcal{I}|$ and logarithms taken to base $q = 757$ for both schemes. 
\begin{table}[h]
     \centering
     \caption{Leakage results.}
     \begin{tabular}{|c |c | c|} 
     \hline
     $|\mathcal{I}|$ & Single-phase I-PCR & Two-phase I-PCR \\ 
     \hline
     0 & 1.1432 & 0.9422 \\
     \hline
     1 & 1.4492 & 0.4200 \\
     \hline
     2 & 1.4492 & 0.0977 \\
     \hline
     3 & 1.1432 & 0.0000 \\
     \hline
     \end{tabular}
     \label{table_leakageValues}
\end{table}

As shown in Table~\ref{table_leakageValues}, the leakage for the two-phase scheme is always less than that for the single-phase scheme. For the single-phase scheme, the leakage values are equal for $|\mathcal{I}|$ and $d-|\mathcal{I}|$. As opposed to the single-phase scheme, the leakage value of the the two-phase scheme decreases as $|\mathcal{I}|$ increases. This is because, the expected size of $\Theta$ reduces, revealing less information on $\mathcal{D}$. The case of $|\mathcal{I}|=d$ is interesting since the user is guaranteed to not find a counterfactual. For the two-phase scheme, the leakage is $0$ since the user knows that $\Theta=\emptyset$ since $x$ is rejected, hence $x\notin \mathcal{D}$. However, this is non-zero for the single-phase scheme.

\bibliographystyle{unsrt}
\bibliography{reference}

\begin{thebibliography}{10}

\bibitem{voigt2017eu}
P.~Voigt and A.~Bussche.
\newblock The {EU} general data protection regulation ({GDPR}).
\newblock {\em A Practical Guide, 1st Ed., Cham: Springer International Publishing}, 10(3152676):10--5555, 2017.

\bibitem{Harvard_discussion_first_counterfactual}
S.~Wachter, B.~Mittelstadt, and C.~Russell.
\newblock Counterfactual explanations without opening the black box: Automated decisions and the gdpr.
\newblock {\em Cybersecurity}, 2017.

\bibitem{nice}
D.~Brughmans, P.~Leyman, and D.~Martens.
\newblock Nice: an algorithm for nearest instance counterfactual explanations.
\newblock {\em Data Mining and Knowledge Discovery}, 38, April 2023.

\bibitem{upadhyayROAR}
S.~Upadhyay, S.~Joshi, and H.~Lakkaraju.
\newblock Towards robust and reliable algorithmic recourse.
\newblock {\em Advances in Neural Information Processing Systems}, 34:16926--16937, 2021.

\bibitem{hammanRobustCF}
F.~Hamman, E.~Noorani, S.~Mishra, D.~Magazzeni, and S.~Dutta.
\newblock Robust counterfactual explanations for neural networks with probabilistic guarantees.
\newblock In {\em International Conference on Machine Learning}, pages 12351--12367. PMLR, 2023.

\bibitem{dutta2022robust}
S.~Dutta, J.~Long, S.~Mishra, C.~Tilli, and D.~Magazzeni.
\newblock Robust counterfactual explanations for tree-based ensembles.
\newblock In {\em International Conference on Machine Learning}, pages 5742--5756. PMLR, 2022.

\bibitem{hammanRobustCFJournal}
F.~Hamman, E.~Noorani, S.~Mishra, D.~Magazzeni, and S.~Dutta.
\newblock Robust algorithmic recourse under model multiplicity with probabilistic guarantees.
\newblock {\em IEEE Journal on Selected Areas in Information Theory}, 5:357--368, 2024.

\bibitem{face}
R.~Poyiadzi, K.~Sokol, R.~Santos-Rodriguez, T.~De Bie, and P.~Flach.
\newblock Face: Feasible and actionable counterfactual explanations.
\newblock Association for Computing Machinery, 2020.

\bibitem{dice}
R.~Mothilal, A.~Sharma, and C.~Tan.
\newblock Explaining machine learning classifiers through diverse counterfactual explanations.
\newblock {\em CoRR}, abs/1905.07697, 2019.

\bibitem{verma2020counterfactual}
S.~Verma, J.~Dickerson, and K.~Hines.
\newblock Counterfactual explanations for machine learning: A review.
\newblock {\em arXiv preprint arXiv:2010.10596}, 2:1, 2020.

\bibitem{guidottiCounterfactualSurvey}
R.~Guidotti.
\newblock Counterfactual explanations and how to find them: literature review and benchmarking.
\newblock {\em Data Mining and Knowledge Discovery}, 38(5), 2024.

\bibitem{mishra2021survey}
S.~Mishra, S.~Dutta, J.~Long, and D.~Magazzeni.
\newblock A survey on the robustness of feature importance and counterfactual explanations.
\newblock {\em arXiv preprint arXiv:2111.00358}, 2021.

\bibitem{pawelczykMembershipInference}
M.~Pawelczyk, H.~Lakkaraju, and S.~Neel.
\newblock On the privacy risks of algorithmic recourse.
\newblock In {\em International Conference on Artificial Intelligence and Statistics}. PMLR, 2023.

\bibitem{yang2022differentially}
F.~Yang, Q.~Feng, K.~Zhou, J.~Chen, and X.~Hu.
\newblock Differentially private counterfactuals via functional mechanism.
\newblock {\em arXiv preprint arXiv:2208.02878}, 2022.

\bibitem{privacy_issue_in_cf}
S.~Goethals, K.~S\"{o}rensen, and D.~Martens.
\newblock The privacy issue of counterfactual explanations: Explanation linkage attacks.
\newblock {\em ACM Trans. Intell. Syst. Technol.}, 14, October 2023.

\bibitem{pentyala2023privacy}
S.~Pentyala, S.~Sharma, S.~Kariyappa, F.~Lecue, and D.~Magazzeni.
\newblock Privacy-preserving algorithmic recourse.
\newblock {\em arXiv preprint arXiv:2311.14137}, 2023.

\bibitem{aivodjiModelExtraction}
U.~A{\"\i}vodji, A.~Bolot, and S.~Gambs.
\newblock Model extraction from counterfactual explanations.
\newblock {\em arXiv preprint arXiv:2009.01884}, 2020.

\bibitem{wangDualCFModelExtraction}
Y.~Wang, H.~Qian, and C.~Miao.
\newblock Dualcf: Efficient model extraction attack from counterfactual explanations.
\newblock In {\em Proceedings of the 2022 ACM Conference on Fairness, Accountability, and Transparency}, 2022.

\bibitem{dissanayakeModelExtraction}
P.~Dissanayake and S.~Dutta.
\newblock Model reconstruction using counterfactual explanations: A perspective from polytope theory.
\newblock In {\em Advances in Neural Information Processing Systems (NeurIPS)}, 2024.

\bibitem{nomeir2024privatecounterfactualretrieval}
M.~Nomeir, P.~Dissanayake, S.~Meel, S.~Dutta, and S.~Ulukus.
\newblock Private counterfactual retrieval.
\newblock {\em arXiv preprint arXiv:2410.13812}, 2024.

\bibitem{chor}
B.~Chor, E.~Kushilevitz, O.~Goldreich, and M.~Sudan.
\newblock Private information retrieval.
\newblock {\em Jour. of the ACM}, 45(6):965--981, November 1998.

\bibitem{c_pir}
H.~Sun and S.~A. Jafar.
\newblock The capacity of private information retrieval.
\newblock {\em IEEE Trans. Info. Theory}, 63(7):4075--4088, July 2017.

\bibitem{c_spir}
H.~Sun and S.~A. Jafar.
\newblock The capacity of symmetric private information retrieval.
\newblock {\em IEEE Trans. Info. Theory}, 65(1):322--329, June 2018.

\bibitem{arbitrarycollusion}
X.~Yao, N.~Liu, and W.~Kang.
\newblock The capacity of private information retrieval under arbitrary collusion patterns for replicated databases.
\newblock {\em IEEE Trans. Info. Theory}, 67(10):6841--6855, July 2021.

\bibitem{banawan_eaves}
K.~Banawan and S.~Ulukus.
\newblock Private information retrieval through wiretap channel {II}: Privacy meets security.
\newblock {\em IEEE Trans. Info. Theory}, 66(7):4129--4149, February 2020.

\bibitem{banawan_multimessage_pir}
K.~Banawan and S.~Ulukus.
\newblock Multi-message private information retrieval: Capacity results and near-optimal schemes.
\newblock {\em IEEE Trans. Info. Theory}, 64(10):6842--6862, April 2018.

\bibitem{banawan_pir_mdscoded}
K.~Banawan and S.~Ulukus.
\newblock The capacity of private information retrieval from coded databases.
\newblock {\em IEEE Trans. Info. Theory}, 64(3):1945--1956, January 2018.

\bibitem{batuhan_hetero}
K.~Banawan, B.~Arasli, Y.-P. Wei, and S.~Ulukus.
\newblock The capacity of private information retrieval from heterogeneous uncoded caching databases.
\newblock {\em IEEE Trans. Info. Theory}, 66(6):3407--3416, June 2020.

\bibitem{byzantine_tpir}
K.~Banawan and S.~Ulukus.
\newblock The capacity of private information retrieval from {B}yzantine and colluding databases.
\newblock {\em IEEE Trans. Info. Theory}, 65(2):1206--1219, September 2018.

\bibitem{C_SETPIR}
Q.~Wang, H.~Sun, and M.~Skoglund.
\newblock The capacity of private information retrieval with eavesdroppers.
\newblock {\em IEEE Transactions on Information Theory}, 65(5):3198--3214, December 2018.

\bibitem{ChaoTian}
C.~Tian, H.~Sun, and J.~Chen.
\newblock Capacity-achieving private information retrieval codes with optimal message size and upload cost.
\newblock {\em IEEE Trans. Info. Theory}, 65(11):7613--7627, November 2019.

\bibitem{codedstorage_adversary_tpir}
L.~Holzbaurand, R.~Freij-Hollanti, and C.~Hollanti.
\newblock On the capacity of private information retrieval from coded, colluding, and adversarial servers.
\newblock In {\em IEEE ITW}, August 2019.

\bibitem{colluding}
H.~Sun and S.~A. Jafar.
\newblock The capacity of robust private information retrieval with colluding databases.
\newblock {\em IEEE Trans. Info. Theory}, 64(4):2361--2370, April 2018.

\bibitem{csa}
Z.~Jia, H.~Sun, and S.~A. Jafar.
\newblock Cross subspace alignment and the asymptotic capacity of {$X$}-secure {$T$}-private information retrieval.
\newblock {\em IEEE Trans. Info. Theory}, 65(9):5783--5798, May 2019.

\bibitem{first_xsecure}
H.~Yang, W.~Shin, and J.~Lee.
\newblock Private information retrieval for secure distributed storage systems.
\newblock {\em IEEE Trans. Info. Foren. Security}, 13(12):2953--2964, May 2018.

\bibitem{wei_banawan_cache_pir}
Y.-P. Wei, K.~Banawan, and S.~Ulukus.
\newblock Fundamental limits of cache-aided private information retrieval with unknown and uncoded prefetching.
\newblock {\em IEEE Trans. Info. Theory}, 65(5):3215--3232, November 2018.

\bibitem{wang_spir}
Z.~Wang and S.~Ulukus.
\newblock Symmetric private information retrieval at the private information retrieval rate.
\newblock {\em IEEE Jour. on Selected Areas in Info. Theory}, 3(2):350--361, June 2022.

\bibitem{ulukusPIRLC}
S.~Ulukus, S.~Avestimehr, M.~Gastpar, S.~A. Jafar, R.~Tandon, and C.~Tian.
\newblock Private retrieval, computing, and learning: Recent progress and future challenges.
\newblock {\em IEEE Journal on Selected Areas in Communications}, 40(3):729--748, March 2022.

\bibitem{uncoded_constrainedstorage_pir}
M.~A. Attia, D.~Kumar, and R.~Tandon.
\newblock The capacity of private information retrieval from uncoded storage constrained databases.
\newblock {\em IEEE Trans. Info. Theory}, 66(11):6617--6634, September 2020.

\bibitem{tspir_mdscoded}
Q.~Wang and M.~Skoglund.
\newblock Symmetric private information retrieval from {MDS} coded distributed storage with non-colluding and colluding servers.
\newblock {\em IEEE Trans. Info. Theory}, 65(8):5160--5175, March 2019.

\bibitem{utah_hetero}
N.~Woolsey, R.~Chen, and M.~Ji.
\newblock Uncoded placement with linear sub-messages for private information retrieval from storage constrained databases.
\newblock {\em IEEE Trans. Commun.}, 68(10):6039--6053, October 2020.

\bibitem{tpir_sideinfo}
Z.~Chen, Z.~Wang, and S.~A. Jafar.
\newblock The capacity of {$T$}-private information retrieval with private side information.
\newblock {\em IEEE Trans. Info. Theory}, 66(8):4761--4773, March 2020.

\bibitem{sun_eaves}
Q.~Wang, H.~Sun, and M.~Skoglund.
\newblock The capacity of private information retrieval with eavesdroppers.
\newblock {\em IEEE Trans. Inf. Theory}, 65(5):3198--3214, December 2018.

\bibitem{semantic_pir}
S.~Vithana, K.~Banawan, and S.~Ulukus.
\newblock Semantic private information retrieval.
\newblock {\em IEEE Trans. Info. Theory}, 68(4):2635--2652, December 2021.

\bibitem{Salim_CodedPIR}
R.~Tajeddine, O.~Gnilke, and S.~El Rouayheb.
\newblock Private information retrieval from {MDS} coded data in distributed storage systems.
\newblock {\em IEEE Trans. Info. Theory}, 64(11):7081--7093, March 2018.

\bibitem{nan_eaves}
J.~Cheng, N.~Liu, W.~Kang, and Y.~Li.
\newblock The capacity of symmetric private information retrieval under arbitrary collusion and eavesdropping patterns.
\newblock {\em IEEE Trans. Info. Foren. Security}, 17:3037--3050, August 2022.

\end{thebibliography}
\end{document}